\documentclass[epj]{svjour}
\usepackage{graphicx}
\begin{document}
\title{Edge states  versus diffusion in  disordered graphene flakes}

\author{Ioannis Kleftogiannis  \and Ilias Amanatidis$^{*}$
}                     
\institute{Department of Physics, University of Ioannina, Ioannina 45110, Greece}
\date{Received: date / Revised version: date}
%
\abstract{We study the localization properties of the wavefunctions in graphene flakes with short range disorder, via the numerical calculation of the Inverse Participation Ratio($IPR$) and it's scaling which provides the fractal dimension $D_{2}$. We show that the edge states which exist at the Dirac point of ballistic graphene (no disorder) with zig-zag edges survive in the presence of weak disorder with wavefunctions localized at the boundaries of the flakes. We argue, that there is a strong interplay between the underlying destructive interference mechanism of the honeycomb lattice of graphene leading to edge states and the diffusive interference mechanism introduced by the short-range disorder. This interplay results in a highly abnormal behavior,  wavefunctions are becoming progressively less localized as the disorder is increased, indicated by the decrease of the average $\langle IPR\rangle$ and the increase of $D_{2}$. We verify, that this abnormal behavior is absent for graphene flakes with armchair edges which do not provide edge states.
} 
\maketitle
\section{Introduction}
\label{intro}
Graphene\cite{Geim,Novoselov} the first 2d metal ever made is an one atom-thick layer of carbon atoms arranged in a honeycomb lattice structure offering large practical advantages over conventional semiconductors, which make it an excellent candidate for replacing silicon in future nanoelectronics.  Most importantly graphene offers a vast field for fundamental theoretical work revealing phenomena such as, relativistic behavior of the electrons at the Fermi level$(E=0)$, known as the Dirac point and topological phenomena like the edge states where the electron current flows along the sample boundaries like in topological insulators.
Edge states have been studied in \cite{fujita,wakaplus,akhemerov} through the theoretical investigation of long stripes of graphene known as nanoribbons \cite{Li,Jiao,Brumfiel,Fuhrer} while they have been experimentally observed in \cite{taoplus} through scanning tunneling microscopy (STM) and spectroscopy (STS) techniques. In \cite{fujita}, it is shown that nanoribbons with the so
called zigzag edges exhibit zero energy edge states with wavefunctions concentrated at the borders of the ribbons. These states are absent for nanoribbons with the other possible type of edge morphology, the armchair edges. So, zigzag nanoribbons exhibit vastly different electronic properties compared to the armchair nanoribbons, at the Fermi level. In principle, the type of edge(zig-zag or armchair) at the boundaries of a graphene system plays a crucial role in its fundamental electronic properties. This argument becomes also apparent  when studying confined graphene structures known as flakes \cite{yamamoto,ezawa,wang,heiskanen1,heiskanen2} which have been experimentally fabricated  in\cite{Geim,ponomarenko,wu,zu,Guttinker,Schnez}. In  \cite{ezawa,heiskanen1,heiskanen2}  it is
shown that graphene flakes with zigzag edges exhibit also edge states which are absent for flakes with armchair edges. For instance, trigonal flakes with zigzag edges provide edge states at the Fermi level while hexagonal flakes with zig-zag edges
give edge states near the Fermi level instead.

Apart from the detailed edge morphology of graphene systems, another important factor that should be taken into account when studying their electronic properties is the presence of disorder which is an inevitable factor as in any mesoscopic material. The main sources of disorder in graphene are the production method (synthesis) and the interaction with the supporting substrate\cite{neto,lewenkopf}. Disorder can appear as lattice distortions like wrinkles,  rippling, or impurities with various degrees of concentration coming from strains in the lattice or charge traps. Disorder is not always an undesirable factor, it can be useful also in applications, for example in spintronic devices through the interaction with the spin\cite{voznyy,oleg} allowing the manipulation of the magnetic properties of disordered graphene systems. Moreover, macroscopic  graphene like lattice structures (honeycomb) with controllable disorder have been shown to be achievable in  \cite{kuhl}  through microwave simulation of the electronic waves. The theoretical  treatment of disordered graphene requires the introduction of different models  like short or long range(smooth)disorder. The type of disorder plays a crucial role on the localization properties of the wavefunctions in graphene. Long range disorder retains the separation between the two so called valleys, centered at the two non equivalent Dirac points of pure  graphene $(E=0)$ at the corners of its hexagonal Brillouin zone, where the relativistic nature of electrons is revealed\cite{neto,lewenkopf,beenacker}. The separation between the two valleys in disordered graphene results
in many interesting phenomena like anti-localization \cite{ando1} or minimum conductance\cite{ando2,wakapcc1,wakapcc2}. On the other hand, short range disorder mixes the two valleys(inter-valley scattering), suppressing the relativistic effects at the Fermi level and resulting in Anderson localization\cite{anderson1,anderson2,nanjing,romer1}. In  \cite{amanatidis,huang,akhmerov,rycerz,amanatidis1} a rough estimation of the localization properties of the wavefunctions can be derived through the study of energy level statistics of disordered graphene flakes, verifying for example the Anderson localization for short range disorder. In general there has been an extensive study of Anderson localization phenomena in graphene.

However, the localization properties of the wavefunctions in the diffusive regime \cite{weinmann,falko} has been much less studied in graphene, especially concerning the cases where the edge states are present. The diffusive regime is defined in systems with short range disorder when the system's length is smaller than the localization length i.e before the onset to localization, where diffusive interference effects are known to dominate the behavior of the wavefunctions. In this regime, the wavefunctions show a chaotic form with the amplitude randomly fluctuating covering the whole system area. Moreover, the random fluctuations of the amplitude follow a multifractal form\cite{falko,Hentschel,spiros,chamon,castillo,subramaniam,evers,romer,kleftogiannis}, a phenomenon that is absent in the localized regime, that is for large scales where Anderson localization is revealed. In essence, this behavior owns it's existence on the finite size of the system, so it is reasonable to use confined structures like flakes for it's investigation.  Specifically, for graphene the study of multifractality when edge states are present has shown interesting effects\cite{kleftogiannis}. So, our main goal in this paper is to investigate the edge states at  the presence of disorder in the diffusive regime, through the study of disordered graphene flakes that provide edge states at the zero disorder limit. Our analysis involves the numerical calculation of the Inverse Participation Ratio ($IPR$) and it's scaling behavior which gives the fractal dimension $D_{2}$ characterizing roughly the volume of a wavefunction.
Both measures combined provide a rough picture of the wavefunction form. Our calculations show evidence of the interplay between two mechanisms: the interference mechanism of the honeycomb lattice of graphene, leading to concentrated  wavefunctions at the borders namely the edge states, and the interference effects leading to diffusion of the wavefunctions in conventional disordered systems. The interaction between these two interference mechanisms has a large impact on the localization properties of the wavefunctions. When edge states are present, we observe a decrease of the average $\langle IPR\rangle$ and an increase of  $D_{2}$ with increasing disorder, implying that the wavefunctions become progressively less localized. This is a highly abnormal behavior compared to the conventional 2d disordered systems, where the wavefunctions become naturally more localized with increasing disorder. We verify that the normal behavior is reproduced for disordered graphene flakes without edge states.

\par The remainder of the paper is organized as follows. In section 2, we introduce our numerical model based on the tight-binding framework along with the short-range  disorder. In section 3 and 4, the numerical results for the various graphene flake shapes at the presence of short-range disorder are presented. We study two different kind of shapes, triangular and hexagonal, with zigzag and armchair edges respectively. We discuss our results and conclude in section 5.

\section{Model}
\label{model}

For our analysis, we use the standard tight binding model for graphene with first nearest neighbor hopping and short-range disorder simulated by a random on-site potential on each lattice site. The model is described by the following Hamiltonian
\begin{equation}
\label{tight}
H= \sum_n \varepsilon_{n}c_{n}^\dagger c_{n}+ \sum_{<n,m>} t( c_{n}^\dagger c_{m} +c_{m}^\dagger c_{n} ) ,
\end{equation}
where $c_n^\dagger$ and $c_n$ are the creation and annihilation operators for spinless fermions, $<n,m>$  denotes nearest neighbors connected with constant hopping element $t$ and $\varepsilon_{n}$ the random on-site potential,  following the box distribution  $P(\varepsilon)=1/w$, in the range $[-w/2,w/2]$ with $w$ denoting the strength of the disorder. Additionally, all energies $E$ are measured in units of the hopping energy $t$, namely $E\equiv E/t$. This type of disorder simulates the existence of impurities in the honeycomb lattice of graphene. It also mixes the two valleys resulting in inter-valley scattering in our problem, which suppresses the relativistic effects at the Fermi energy while it breaks also the chiral symmetry of the graphene lattice. We focus our study on graphene flakes with specific shapes that are known to exhibit edge states at the limit of zero disorder. These include shapes studied in  \cite{ezawa,heiskanen1} like the trigonal and the hexagonal, with zig-zag edges. Also, we extend our study on the same shape types but with armchair edges, investigated also in \cite{ezawa,heiskanen1} where the edge states are absent.  We consider flakes that consist between a few hundred to a few thousand atoms ,as the ones that have been studied experimentally in \cite{ezawa,ponomarenko}.

Diagonalizing the Hamiltonians for each graphene flake we derive the wavefunction amplitudes $\Psi_{i}$ at each lattice site i for energy E, which allow us to calculate the inverse participation ratio $(IPR)$ given by $IPR(E) = \sum_i|\Psi_{i}(E)|^{4} $. The participation ratio gives the information about the degree in which every site of the lattice is participating in the wavefunction. With its inverse we can get a rough estimation of the localization properties. In general, wavefunctions with densely distributed amplitude for example a diffusive or a ballistic wavefunction, will give values of $IPR$ close to zero while localized wavefunctions will give $IPR$ values close to one. In addition, the scaling of $IPR$ provides the fractal dimension $D_{2}$ of a wavefunction through the relation $IPR\sim L^{-D_{2}}$ and is a characteristic measure of the volume it roughly occupies.  For instance, when analyzing a two dimensional system, $D_{2}=2$ when the wavefunction amplitude is equally distributed on the whole lattice while $D_{2}=0$ for a localized wavefunction. In general, $D_{2}$ belongs to a spectrum of fractal dimensions characterizing objects known as multifractals. Non integer values of $D_{2}$ imply fractal wavefunctions\cite{spiros,chamon,castillo,romer,kleftogiannis} that are not either extended nor localized.

The value of $D_{2}$ gives a good estimation of the volume a wavefunction is occupying although it does not uniquely identify it's form. For this reason, we illustrate the wave-function probability amplitude $|\Psi_{i}(E)|^{2}$ for each case that we are studying. With these three measures the $IPR$, $D_{2}$ and the wavefunction probability, we can get a fairly complete picture of the wavefunctions in disordered graphene flakes.

\section{Triangular graphene flakes}
\label{triang_graph_flakes}

\begin{figure}
         \centering
          \includegraphics[width=0.9\columnwidth,clip=true]{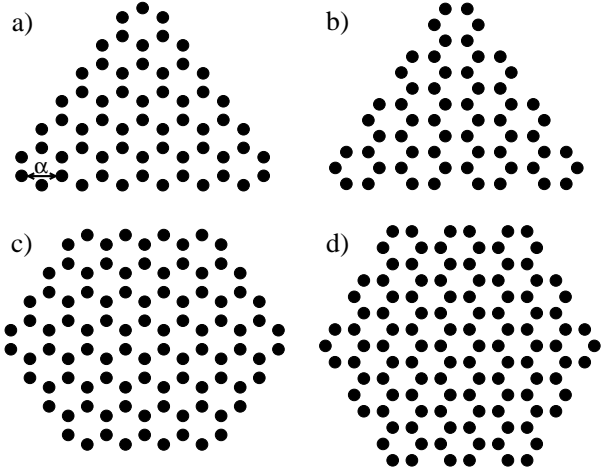}
         \caption{The flake shapes we study, triangular and hexagonal. a) Equilateral triangular graphene flake with zigzag edges and b)armchair edges. c) Hexagonal flake with zigzag and d) armchair edges. We choose the side length L of each shape, measured in units of the lattice constant $\alpha$, to characterize the flake size (linear length scale). The flakes with zigzag edges provide edge states at the zero disorder limit in contrast to the flakes with armchair edges.}
         \label{fig1}
\end{figure}

In this section, we examine the wavefunction localization properties of disordered graphene flakes of trigonal shape with two different types of edges, zigzag and armchair (Fig. 1), via the analysis of $IPR$  and its scaling  which gives the fractal dimension $D_{2}$. One example of a trigonal graphene flake with zigzag edges can be seen in Fig.1(a). We characterize the size of the flake by the base length of the triangular shape L in units of the lattice constant $\alpha=2.42{\AA}$, for example the length is $L=6\alpha$  in Fig. 1(a). In the case of zero disorder the zig-zag triangle exhibits zero energy edge states as discussed in \cite{ezawa,heiskanen1}, with wavefunctions concentrated on the zigzag edges(see Fig. 3(a)). In Fig. 2, we plot the average value $\langle IPR\rangle$ versus the energy E near the Fermi energy($E=0$)  for a flake with $L=44$ consisting of 2113 atoms and for different disorder strengths $w=0.5,1,1.5$. The number of disorder realizations is 5000. Orange color in the background represents the individual values of $IPR$  for $w=0.5$ used to obtain the corresponding curve of $\langle IPR\rangle$. The  $\langle IPR\rangle$ for w=0 can be also seen as large individual green colored dots, separated by large gaps. The green dot corresponding to the  lowest energy for w=0 is the average $\langle IPR\rangle$ over 43 edge states with energy $E=10^{-16}$. The number of these edge states is significantly larger than the number of extended states(6 states) that lie higher in the energy spectrum. At the presence of disorder both the edge states and the extended states disperse creating two separated energy areas characterized by vastly different values of $IPR$, something that is especially evident for the weaker disorder w=0.5.  For the  low energy regime with the larger $IPR$ values, approximately until $E\sim0.1$ clearly the $IPR$ is in average decreasing with increasing disorder strength. This is also evident in the inset of  Fig. 2 where we show the corresponding values of $IPR$. The points concentrate progressively lower with increasing disorder resulting in lower values of $\langle IPR\rangle$ as we have seen in the main figure. Keeping in mind the two trivial limits of $IPR$, $IPR=0$ for extended wavefunctions(ballistic) and $IPR=1$ for localized wavefunctions, the overall behavior of $IPR$ implies that the low energy wavefunctions of the zigzag triangle with disorder are becoming in average less localized as the disorder is increased. This behavior is highly abnormal and is absent in normal disordered materials, for instance a square lattice or a chain, where the wavefunctions become naturally more localized with increasing disorder, with the $IPR$ increasing in average. We have verified that the abnormal behavior starts approximately at w=0.25. From w=0 to w=0.25 a normal behavior occurs with $IPR$ increasing. In Fig.2, we can also observe an abrupt change of $IPR$ starting at  $E\sim0.1$ resulting in large fluctuations until $E\sim0.17$. This is especially evident for $w=0.5$, in both the curve of $\langle IPR\rangle$ and the individual values of $IPR$(orange color). Below  $E\sim0.1$,  we can distinguish a whole area of points with values of $IPR$ in average much higher than the values corresponding to energies in the interval $E\sim0.17-0.25$.

The energy area $E\sim0-0.1$ consists mainly of wavefunctions that have their amplitude concentrated at the edges of the trigonal flake as shown in figures 3(b),3(c) where the wavefunction probability is plotted for $E\sim0.07$ and different strengths of disorder. Comparing the edge state in Fig. 3(a) for w=0 with the wavefunction in Fig. 3(b) we can see that weak disorder $(w=0.5)$ localizes the wavefunction along a random area on the border of the flake acting in this way as a pertubation on the  zero disorder limit studied in \cite{heiskanen1}, where the edge state amplitude spreads almost periodically along the whole border. So, we can say that the edge states survive for weak disorder in the sense that the amplitude remains mostly concentrated at the border of the flake. The abnormal behavior that we distinguished through the analysis of $IPR$ in the energy area $E\sim0-0.1$ can be understood by looking at figures 3(c) and 3(d) where we plot the wavefunction probability for stronger disorder $w=1.5$ and $w=5$. In Fig. 3(c) the amplitude although still mostly localized along the border, has started extending across it while it also penetrates slightly the flake,  resulting in less localized wavefunctions and in lower $IPR$ values as seen in Fig. 2. Even larger disorder $(w=5)$ in Fig. 3(d) tends to localize the wavefunction inside the flake instead of the edges. For sufficiently strong disorder, the destructive interference mechanism of the honeycomb lattice that leads to edge states in graphene is completely destroyed by the interference mechanism coming from the short-range disorder, leading to Anderson localization with the wavefunctions becoming completely localized in the interior(bulk) of the flake instead of the edges. In essence, the on-site short-range disorder destroys the special topology of the honeycomb lattice that favors the creation of edge states. However,the effects of the edge states are not immediately washed out but only in the large disorder limit. For weaker disorder the abnormal behavior we analyzed occurs, coming from the interplay between the interference mechanisms of the diffusion and the edge states resulting in less localized wavefunctions with increasing disorder. Apart from this interplay another factor that plays a role in the abnormal behavior we obtain is the progressive mixing of the edge states with the extended states as the disorder is increased, which becomes more important for strong disorder. We have considered low disorder strength values $(w<2)$ in the study of $IPR$ in order to minimize the effect of this mechanism. For energies above $E\sim0.17$ in Fig. 2 the localization properties of the wavefunctions change drastically, spreading along the whole flake, indicated by $IPR$ obtaining much lower values than for $E\sim0-0.1$. The transition from edge states to these extended states creates the large fluctuations in the energy interval $E\sim 0.1 - 0.17$.

\begin{figure}
         \centering
            \includegraphics[width=0.9\columnwidth,clip=true]{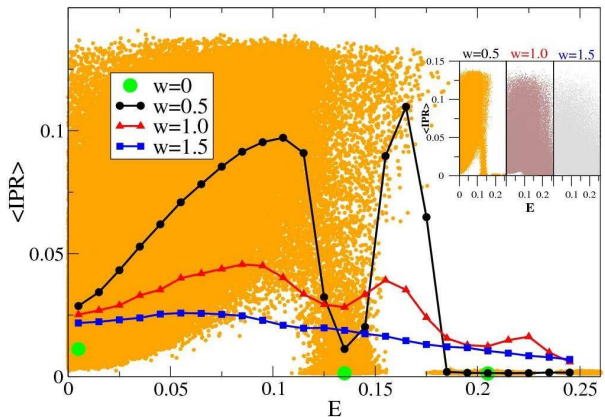}

         \caption{(color online). The average value of the Inverse Participation Ratio ($\langle IPR \rangle$) versus the energy E for a trigonal graphene flake with zigzag edges with $L=44$(2113 sites), for disorder strengths $w=0.5, 1, 1.5$ and 5000 realizations, along with the $w=0$ case(green dots). The orange points in the background represent the individual values of $IPR$ for w=0.5 while the cases for the other disorder strengths are shown in the inset. $\langle IPR \rangle$ decreases with increasing disorder, implying less localized wavefunctions, visible also in the inset, where $IPR$ concentrates on lower values with increasing disorder.  There is an abrupt change of $\langle IPR \rangle$  at $E\sim0.1$ evident also in the background(orange points), coming from the transition from edge states concentrated at low energies to extended states for higher energies.}
         \label{fig2}
\end{figure}

\begin{figure}
\begin{center}$
\begin{array}{cc}
\includegraphics[width=0.35\columnwidth,angle=270]{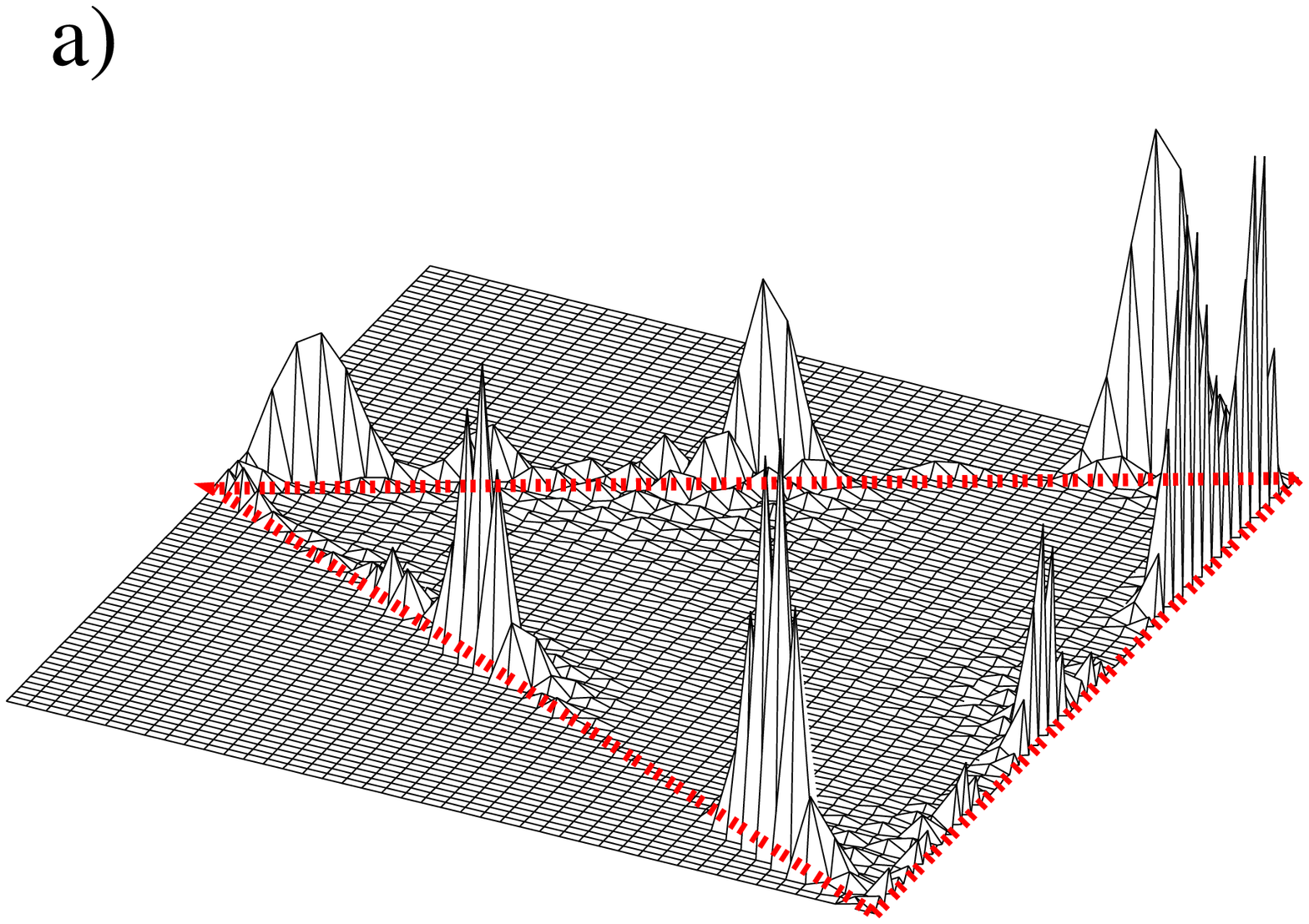}
\includegraphics[width=0.35\columnwidth,angle=270]{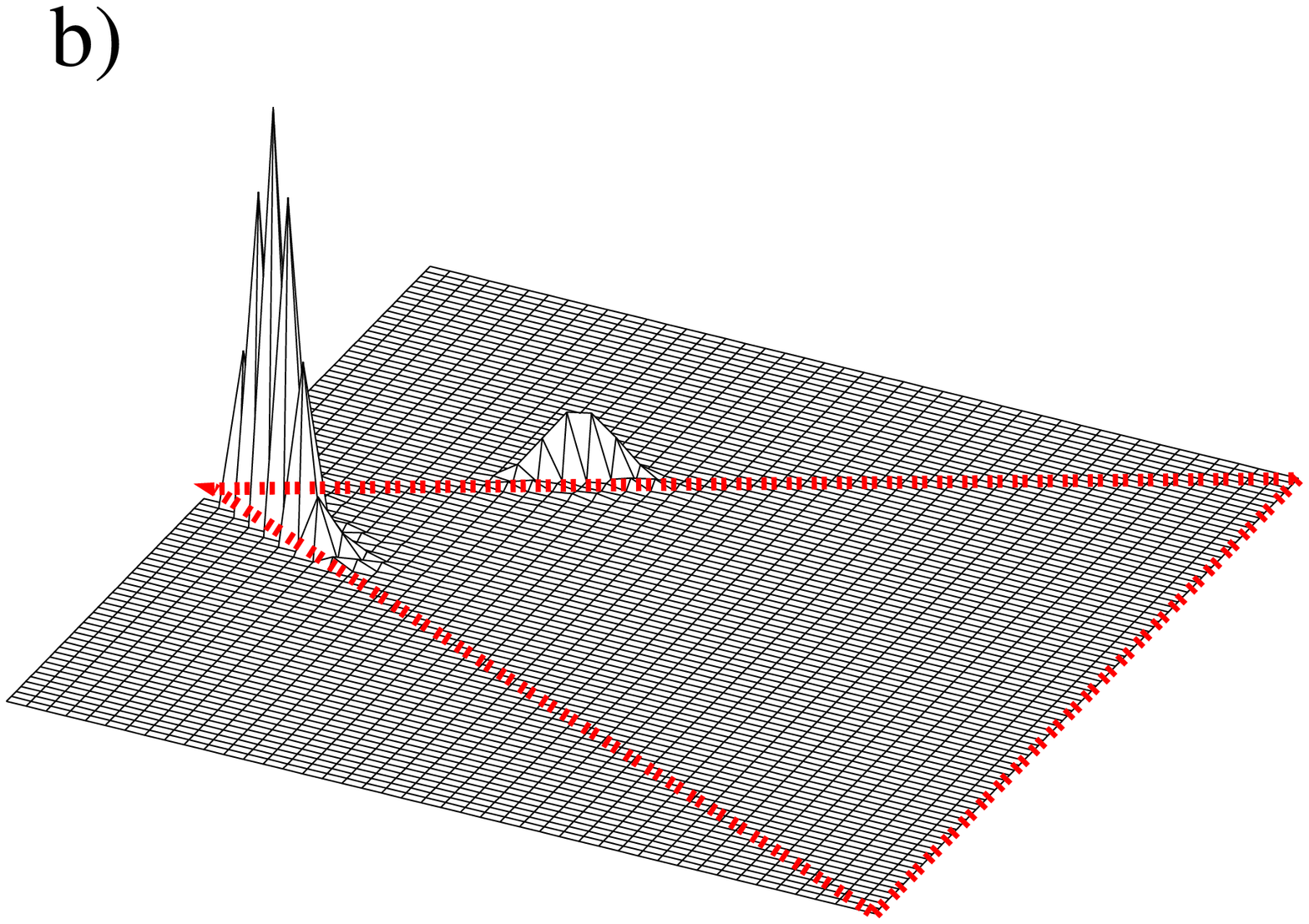}\\
\includegraphics[width=0.35\columnwidth,angle=270]{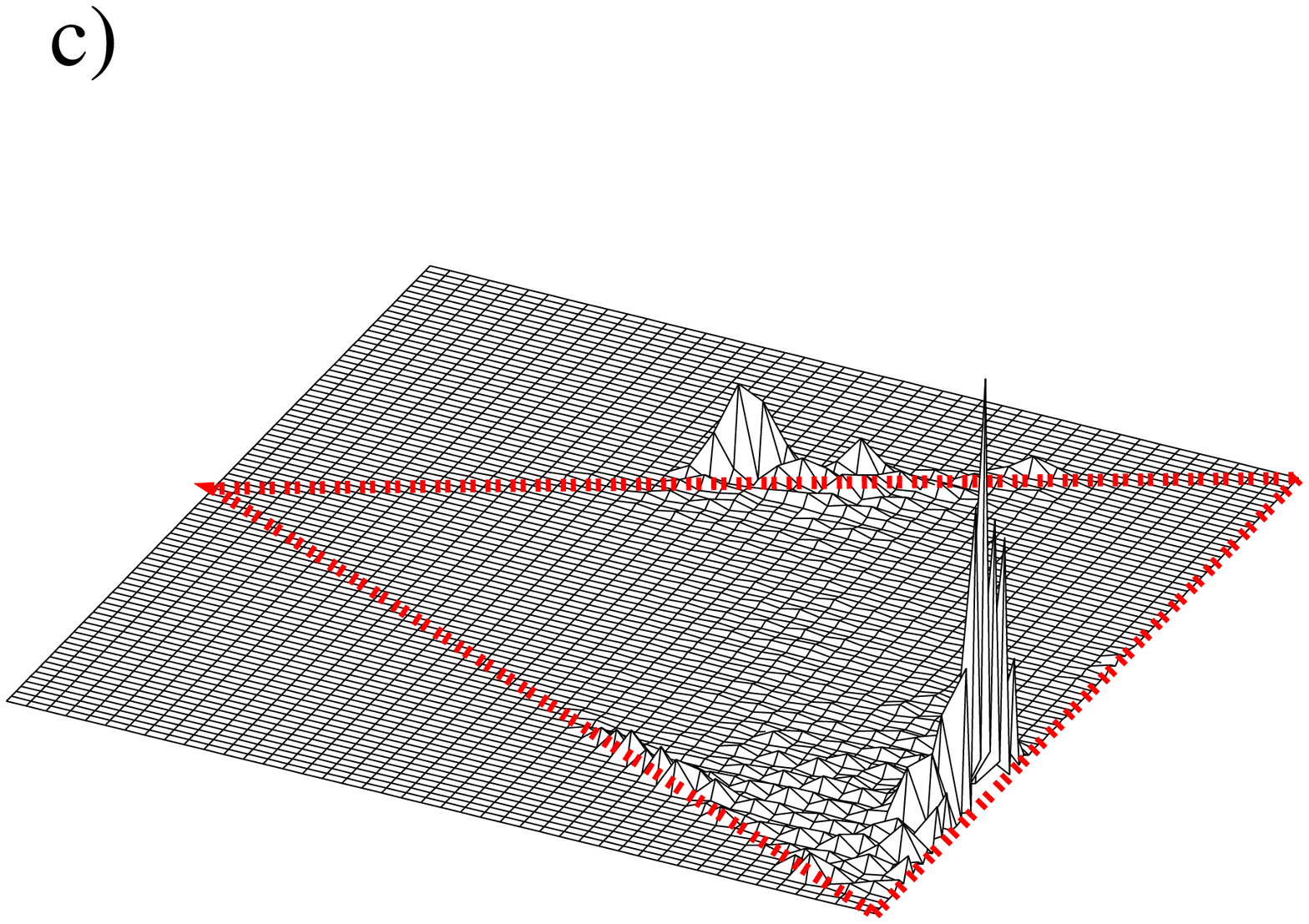}
\includegraphics[width=0.35\columnwidth,angle=270]{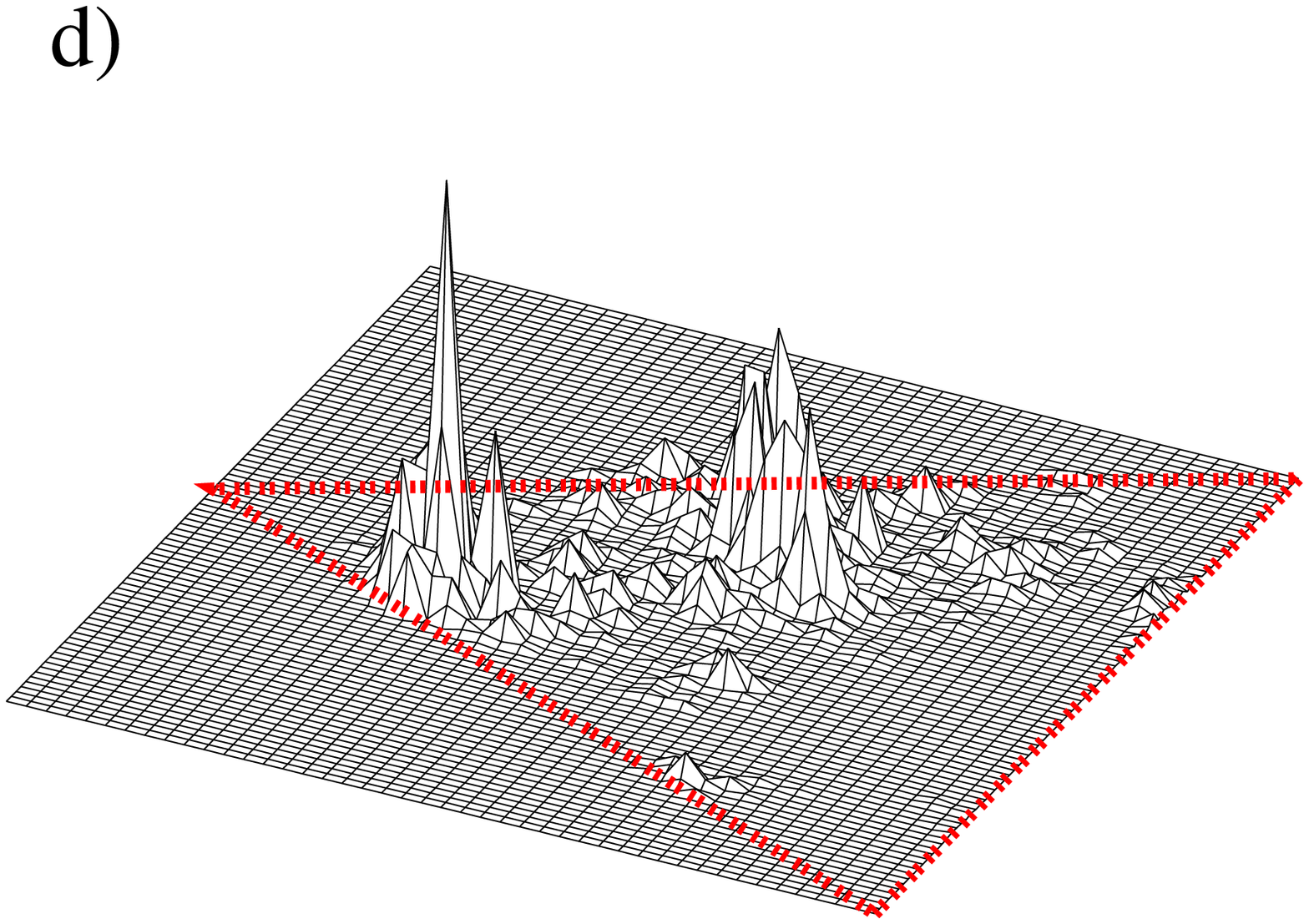}
\end{array}$
\end{center}\caption{(color online). The wavefunction probability amplitude $|\Psi|^{2}$ for a flake with  L=$44$(2113 sites) for different disorder strengths $w=0.5,1.5,5$ at energy  $E\sim0.07$(one disorder realization) along with a wavefunction for $w=0$ at $E=10^{-6}$. (a)For $w=0$ the amplitude is concentrated on the edges of the flake(edge state). (b)For $w=0.5$ the wavefunction becomes localized along the border. (c)For $w=1.5$, the amplitude penetrates slightly the flake, despite being mostly concentrated on the border, showing also abrupt fluctuations. (d)The wavefunction for large disorder strength $w=5$ concentrates in the flake's bulk. There is no sign of the edge states in this case, the wavefunction becomes localized inside the flake instead.
}
\label{fig3}
\end{figure}

\begin{figure}
          \centering
         \includegraphics[width=0.9\columnwidth,clip=true]{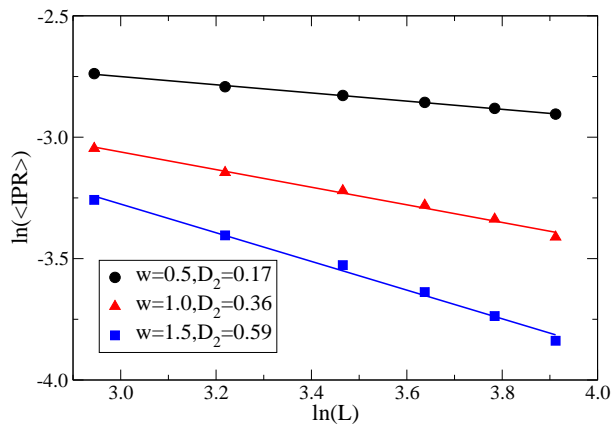}
         \caption{(color online). The scaling of $IPR$ for a trigonal zig-zag flake for different disorder strengths, averaged over energies in the interval $[0,0.1]$  and over 5000 realizations of the disorder. The slope of $\textrm{ln}(\langle IPR\rangle)$ versus $\textrm{ln}(L)$ gives the fractal dimension $D_{2}$ which allows the estimation of the wavefunction volume. It is clearly increased with increasing disorder while the points for $IPR$ for small disorder strengths lie above the corresponding points for larger disorder in agreement with the behavior demonstrated in Fig. 2 considering the average behavior of $IPR$.}
         \label{Fig4}
\end{figure}

We continue our analysis by studying the scaling of $IPR$ from which we can derive the fractal dimension $D_{2}$ providing a rough estimation of the wavefunction's volume. In Fig. 4, we show $D_{2}$  for different strengths of disorder with the values of the average $\langle IPR\rangle$ for each size being over 5000 disorder realizations and energies inside the window  $[0,0.1]$, where the edge states lie approximately according to our previous analysis. Since $IPR\sim L^{-D_{2}}$, we plot $ln(\langle IPR\rangle)$ versus $ln(L)$ in order to get the exact value of $D_{2}$. We can observe that the slope of each curve representing $D_{2}$ increases with increasing disorder strength, implying  that the volume occupied by the corresponding wavefunctions increases also. Moreover, $ln(\langle IPR\rangle)$ and consequently  $\langle IPR\rangle$ for each individual size averaged over the energies and realizations decreases, in agreement with the results obtained in Fig. 2 for the curves of  $\langle IPR\rangle$ versus E. Additionally, $D_{2}$ obtains non integer values implying multifractality inside the chosen energy window, evident from the abrupt fluctuations of the amplitude in Fig. 3.  The values of $D_{2}$  below one are reasonable considering the wavefunctions  for $w=0.5, 1.5$ in figures 3(b),3(c) being mostly concentrated along the border of the trigonal flake extending slightly inside. The overall behavior of  $D_{2}$  versus the disorder strength for the graphene triangle with zigzag edges can be seen in Fig. 14. We should clarify that we have restricted
our analysis of $D_{2}$ on the diffussive regime, for larger flake sizes Anderson localization takes place in all cases giving zero $D_{2}$. So, we have found that although weak disorder preserves the edge states in trigonal flakes with zig-zag edges, its increase results in a highly abnormal behavior, the edge states become progressively less localized, extending inside the flakes.

We contrast this result to the behavior of the $IPR$ and $D_{2}$ observed for trigonal flakes with armchair edges shown in Fig. 1(b), for which edge states are absent for zero disorder according to \cite{ezawa,heiskanen1}. $\langle IPR\rangle$  versus the energy E for  $L=44.5$(2106 atoms)and for different disorder strengths can be seen in Fig.5 along with their respective distributions of $IPR$ in the background(orange points-$w=0.5$) and inside the inset(brown points-$w=1.0$, grey points-$w=1.5$). Also, the $<IPR>$ for w=0 can be seen as large green dots. Clearly, as the disorder is increased,  $\langle IPR\rangle$ increases inside the whole energy window $[E=0-0.26]$ , which is evident also in the insets where the individual points of $IPR$  concentrate progressively in higher values with increasing disorder. The gaps at specific energies for $w=0.5$ are a consequence of the respective gaps appearing in the energy spectrum for zero disorder as seen from the $<IPR>$ for $w=0$. Apart from these gaps in Fig. 5, $IPR$ behaves smoothly with the energy in contrast to the trigonal flake (Fig. 2) where we observed two regions with vastly different values of $IPR$. In Fig. 6, we show a characteristic example of a wavefunction lying inside the energy window of Fig. 5 at $E\sim0.07$. The amplitude fluctuates wildly, randomly spreading on the whole flake, a common picture of a diffusive wavefunction.

\begin{figure}
          \centering
                 \includegraphics[width=0.9\columnwidth,clip=true]{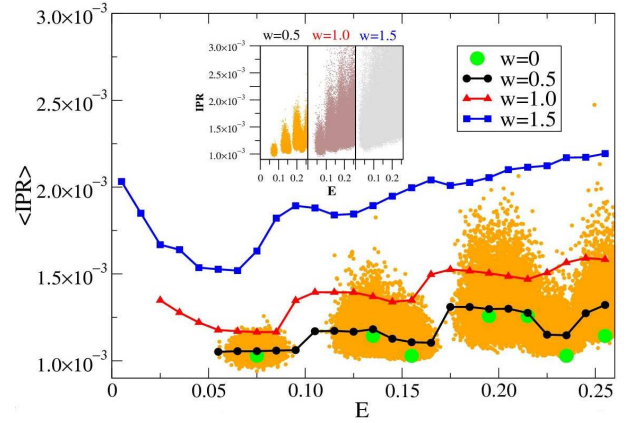}
          \caption{$\langle IPR \rangle$ versus the energy for a trigonal graphene flake with armchair edges of size $L=44.5$(2106sites) for disorder strengths $w=0.5, 1, 1.5$ and $5000$ realizations, along with the $w=0$ case(green dots). The orange points in the background are the individual values of $IPR$ for $w=0.5$, with the other cases shown in the inset. $IPR$ increases in average with increasing disorder, implying more localized wavefunctions, a behavior observed in normal disordered systems like the square lattice. The gaps appearing at the distribution of $IPR$ for $w=0.5$ (orange points) come from the respective gaps present for $w=0$.  As the disorder is increased the gaps are disappearing, as shown in the inset.
          }
          \label{Fig5}
\end{figure}

\begin{figure}
\centering
\includegraphics[width=0.5\columnwidth,angle=270]{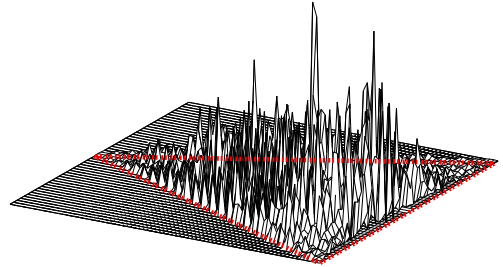}
\caption{ The wavefunction probability amplitude for an armchair trigonal flake of size $L=44.5$(2106 sites) for strength of disorder $w=1.5$ at $E\sim0.07$. The amplitude spreads randomly all over the flake, this is a common picture of a diffusive wavefunction encountered in the diffusive regime of normal 2d systems.
}
\label{fig6}
\end{figure}

The results for the fractal dimension $D_{2}$  can be seen in Fig. 7 for different strengths of disorder.
The value of $IPR$ for each size is calculated as an average over different realizations of the disorder and over the energy window $[0,0.2]$. In contrast to the trigonal flakes with zigzag edges, $D_{2}$  decreases with increasing disorder, as for normal 2d disordered systems where the wavefunctions occupy progressively less volume, becoming less dense, as more disorder is introduced. The overall behavior of $D_{2}$ versus the disorder strength can be seen in Fig. 14. We should also remark that for sufficiently large flakes $D_{2}$ goes to zero because of Anderson localization. So, the trigonal flakes with armchair edges at the presence of short range disorder do not exhibit the abnormal behavior we encountered in the disordered trigonal flakes with zigzag edges. Instead, they behave as normal disordered metals with their corresponding wavefunctions becoming more localized with increasing disorder.

\section{Hexagonal graphene flakes}
\label{hex_graph_flakes}
We now extend our analysis on hexagonal graphene flakes with zigzag edges which provide edge states at the zero disorder limit(see Fig. 9(a)) as in the case of the corresponding trigonal flakes\cite{ezawa,heiskanen1} concentrated near the Fermi energy. Again, we are interested in the effect of the disorder on the edge states obtained through the scaling analysis of $IPR$. The overall behavior of $IPR$ versus the energy can be seen in Fig. 8 for a hexagonal flake with $L=18$ consisting of 1944 sites. The behavior is similar to that of the zig-zag triangle. For low energies below $E\sim0.1$ where the edge states are concentrated(see figures 9(b),9(c)), $<IPR>$ and $IPR$ are decreasing with increasing disorder meaning that the wavefunctions are becoming progressively less localized as indicated by the wavefunction form in Fig. 9(c) compared to 9(b). This abnormal behavior starts approximately from w=0.25 as in the zig-zag triangles. $IPR$ changes drastically above $E\sim0.1$ where extended states start to appear, obtaining much lower values despite the fact that the transition from the edge states to the extended states is smoother in this case than it is for the zigzag triangle. This is because of the denser energy spectrum compared to the triangular flakes, evident from the comparison of $\langle IPR\rangle$ for w=0(green points) between Fig. 2 and Fig. 8. Sufficiently strong disorder localizes the wavefunction inside the flake(see Fig. 9(d)). As in the case of the zigzag trigonal flakes, $D_{2}$ in Fig. 10 is increased as we increase the disorder (wavefunction volume increases), while it's non-integer values imply multifractality in agreement with the  abrupt fluctuations of the amplitude in Fig. 9. It is clear that for weak disorder the edge states survive also in the case of the hexagonal flakes with zigzag edges and result in less localized states as the disorder is increased. So, the abnormal behavior we pointed out for the trigonal flakes with zigzag edges exists also in the  case of hexagonal flakes with zigzag edges. We  conclude that this effect is independent of the overall flake shape and is related with the existence of edge states, governed by the detailed edge structure on the borders of the graphene flakes.

\begin{figure}
          \centering
           \includegraphics[width=0.9\columnwidth,clip=true]{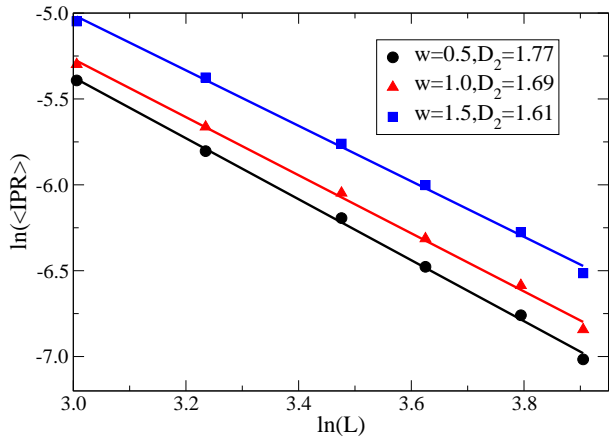}
         \caption{The scaling of $IPR$  for a trigonal armchair flake for different disorder strengths, averaged over the energy interval $[0,0.2]$ and over 5000 realizations. The calculated slope $D_{2}$ characterizing the wavefunction volume is clearly decreased with increasing disorder while the points along the curves increase their values, in agreement with the average behavior of $IPR$ observed in Fig. 4.}
         \label{Fig7}
\end{figure}

\begin{figure}
          \centering
                   \includegraphics[width=0.9\columnwidth,clip=true]{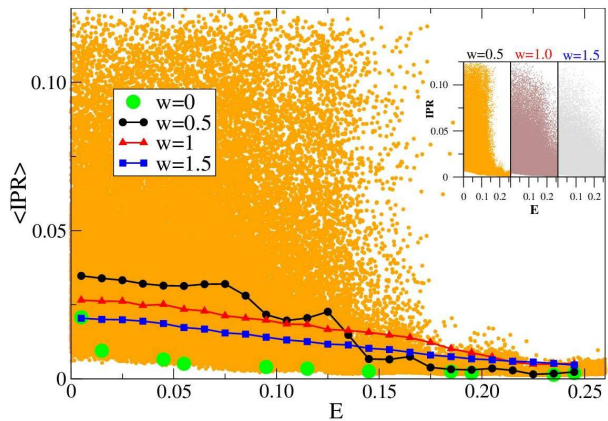}
          \caption{ $\langle IPR \rangle$ vs. E for a hexagonal flake with zigzag edges of size $L=18$(1944 sites) for different disorder strengths and 5000 realizations, along with the case w=0(green points) and the individual values of $IPR$ in the background (orange points-$w=0.5$) and the inset. $\langle IPR \rangle$ decreases with increasing disorder in agreement with the behavior of the $IPR$ in the inset, implying in overall less localized wavefunctions. Around $E\sim0.15$ there is a transition from edge states with high $IPR$ values to extended states with low $IPR$ values. The overall behavior is qualitatively similar to that of the trigonal flake with zigzag edge.
          }
          \label{Fig8}
\end{figure}

\begin{figure}
\begin{center}$
\begin{array}{cc}
\includegraphics[width=0.35\columnwidth,angle=270]{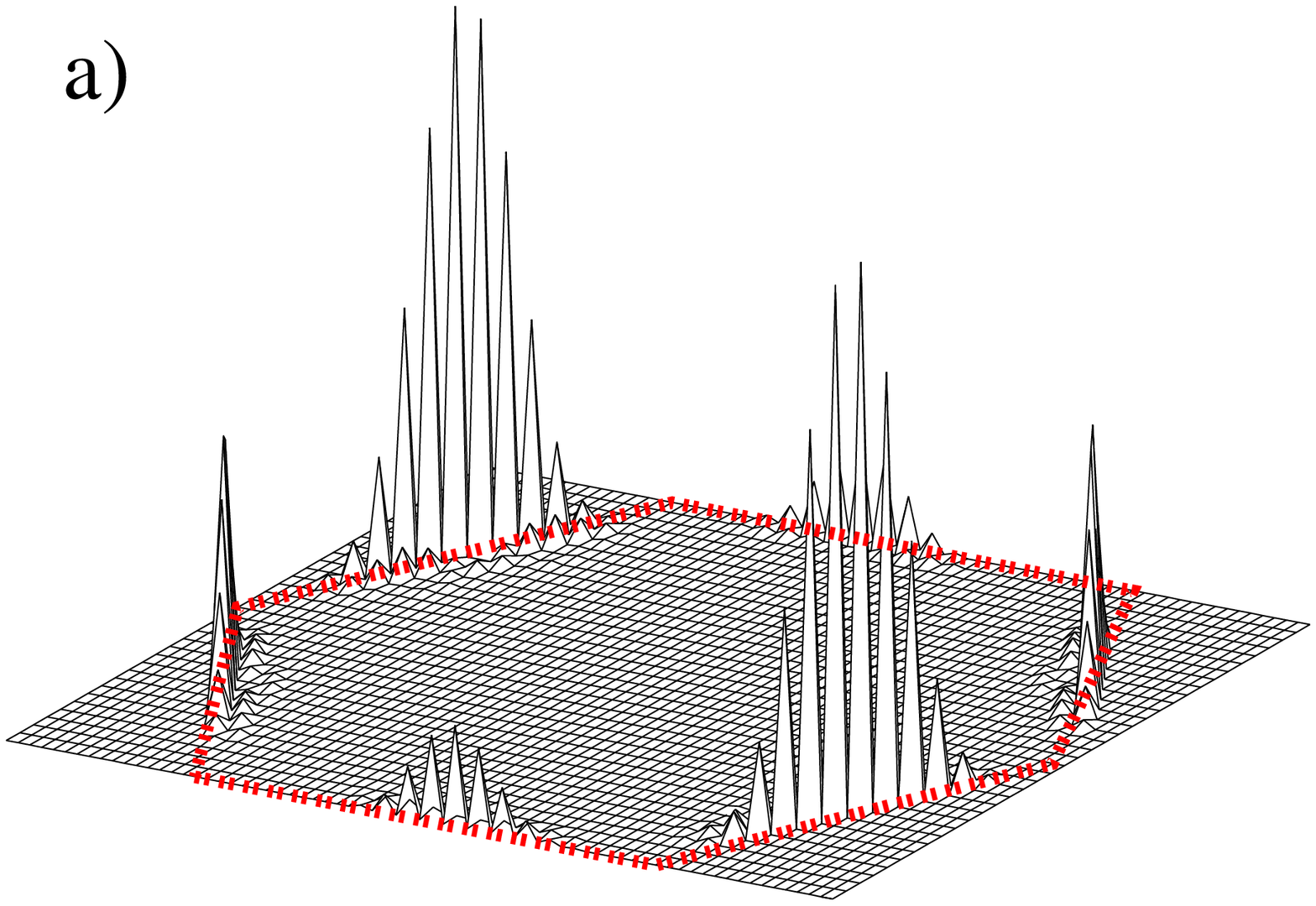}
\includegraphics[width=0.35\columnwidth,angle=270]{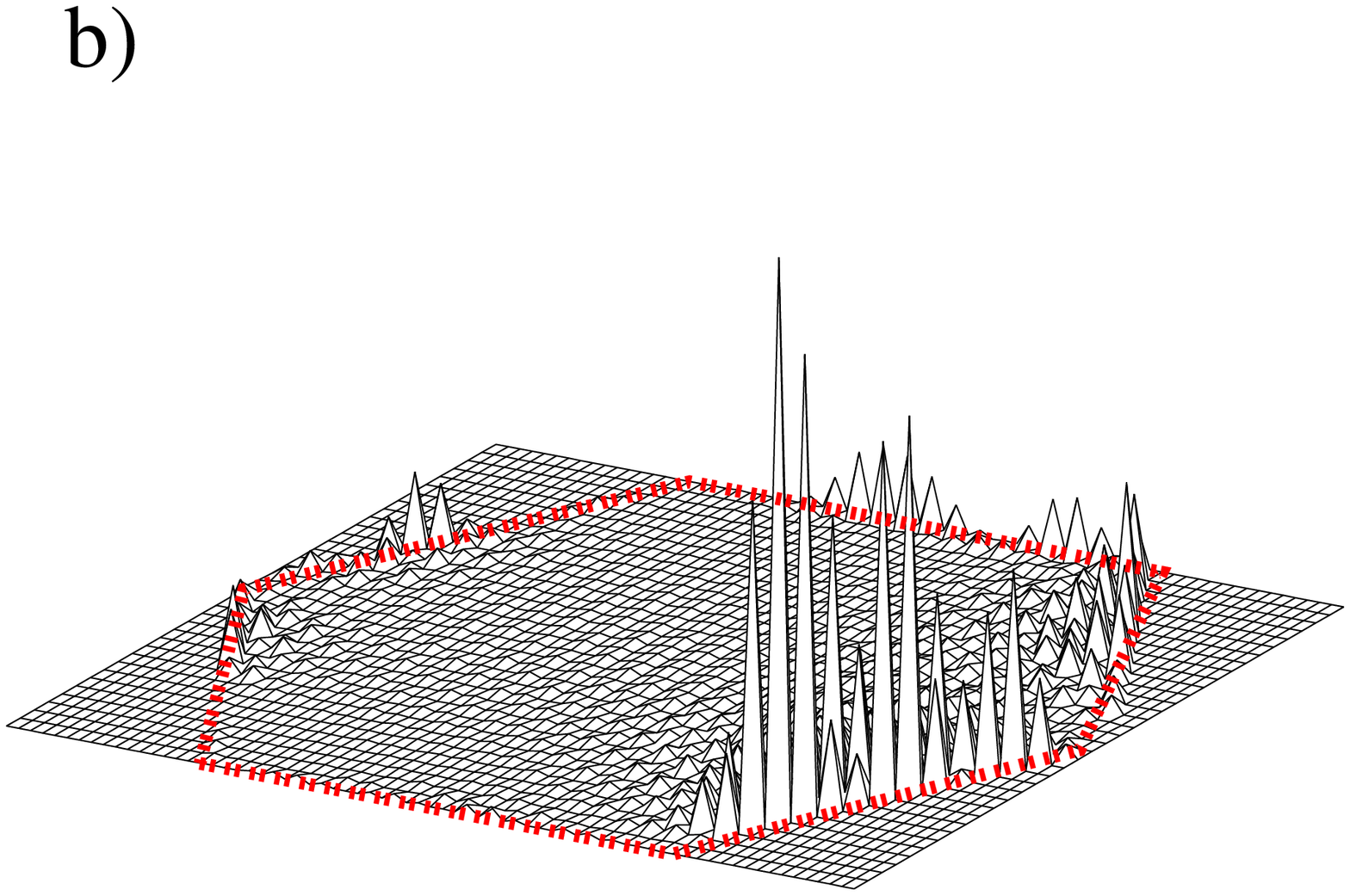}\\
\includegraphics[width=0.35\columnwidth,angle=270]{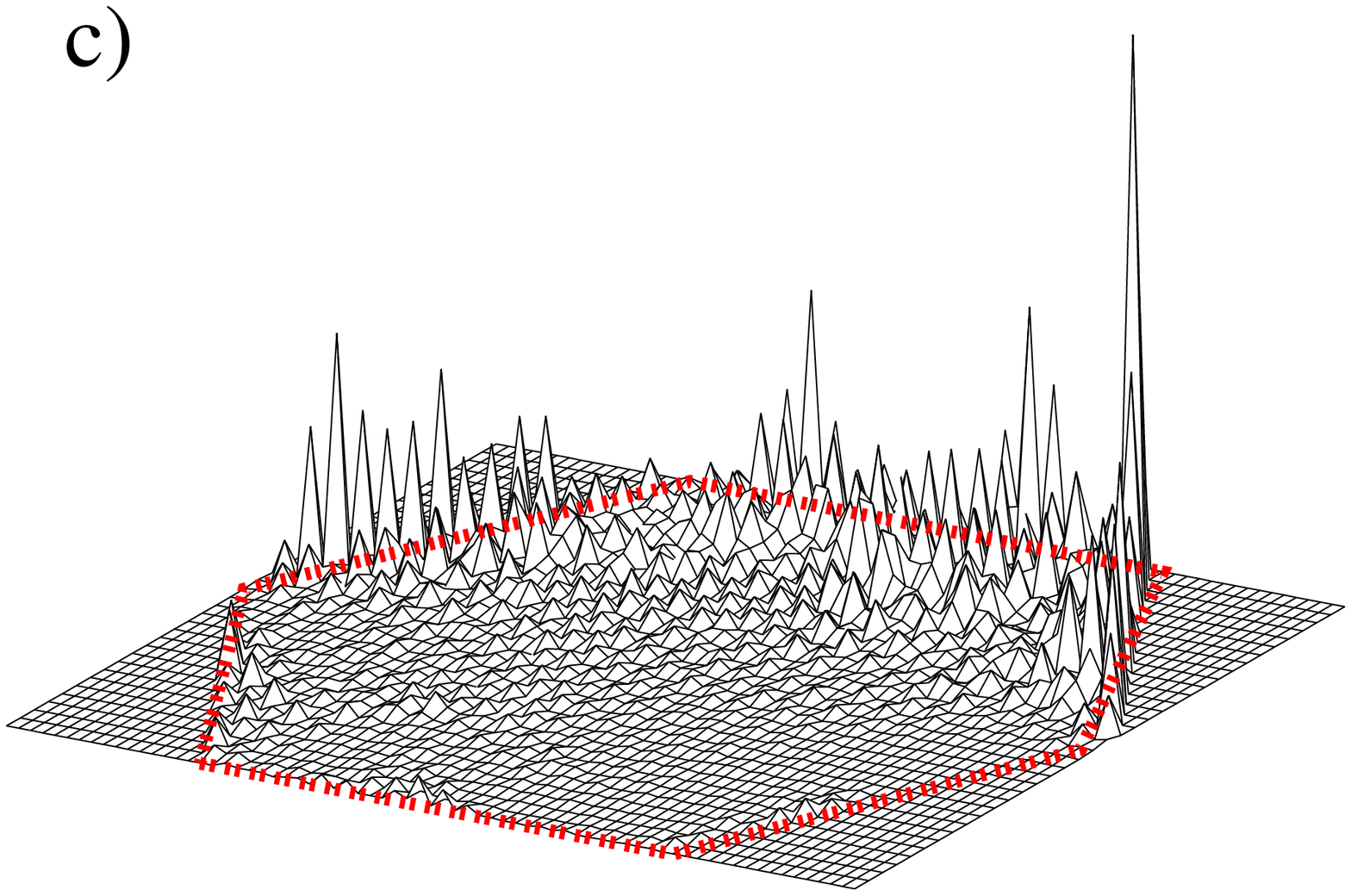}
\includegraphics[width=0.35\columnwidth,angle=270]{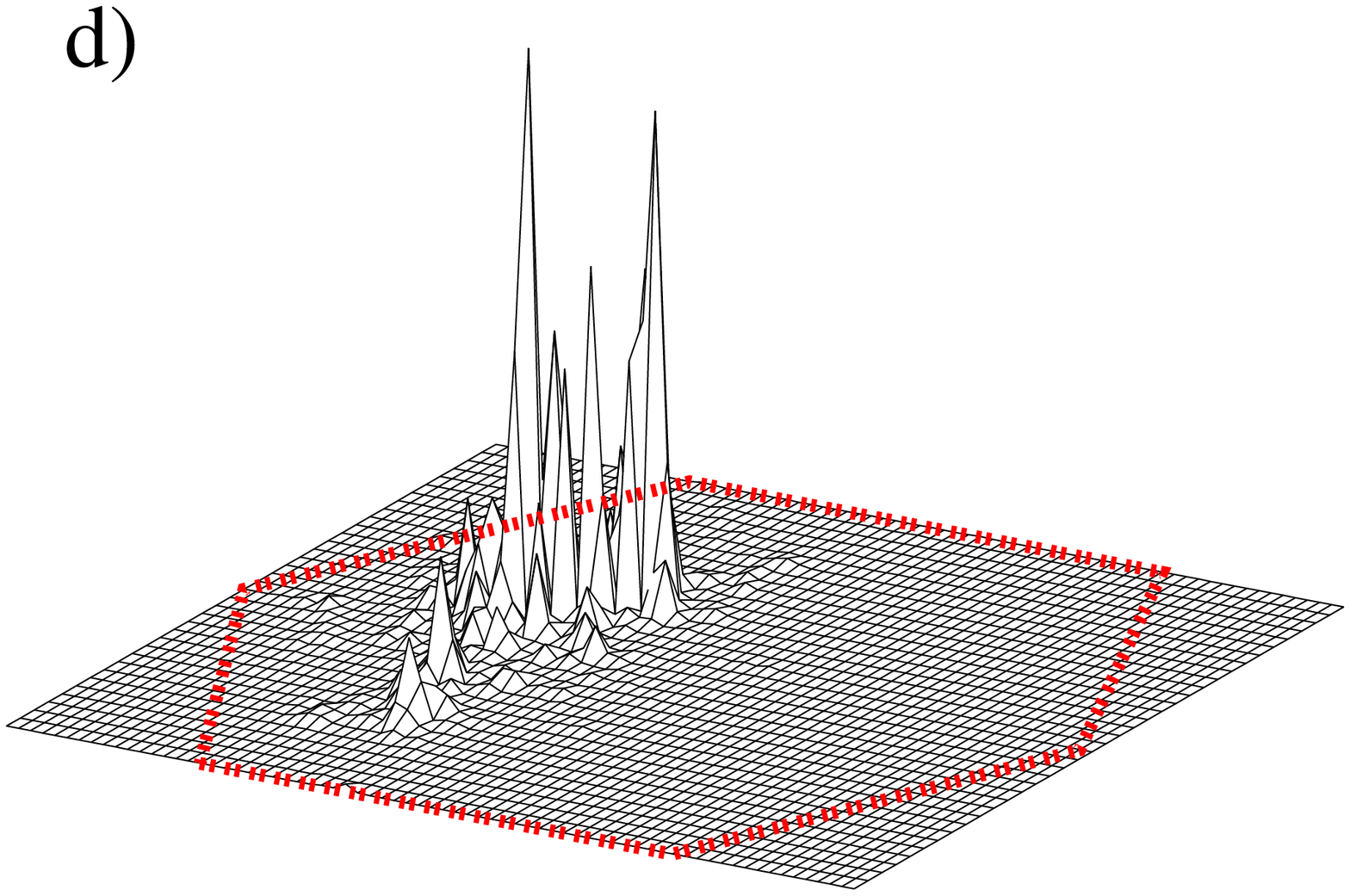}
\end{array}$
\end{center}
\caption{
The wavefunction probability amplitude for a flake of size $L=184$(1944 sites) for  different disorder strengths $w=0.5,1.5,5$ at $E\sim0.07$ along with the $w=0$ case at $E\sim0.0005$ . (a)The amplitude for $w=0$ clearly extends along the border of the flake. (b)For $w=0.5$, the amplitude remains concentrated on the border although localized in a specific area. (c)For $w=1.5$, the amplitude clearly starts penetrating the flake. (d)For strong disorder $w=5$, the wavefunction becomes localized in the bulk. The overall behavior is similar to the trigonal flakes with zigzag edges.}
\label{Fig9}
\end{figure}

\begin{figure}
          \centering
                   \includegraphics[width=0.9\columnwidth,clip=true]{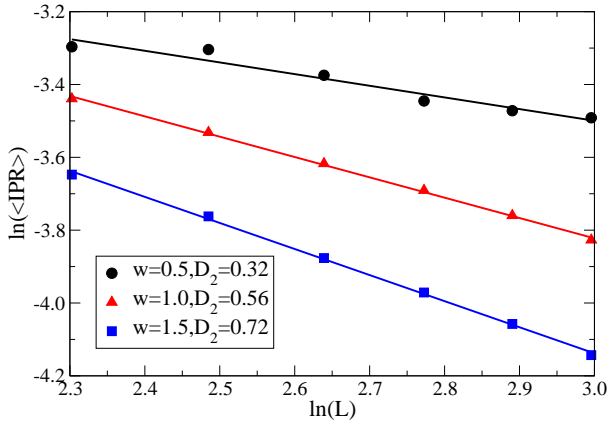}
          \caption{The scaling of $IPR$ for a hexagonal zig-zag flake for increasing disorder strengths averaged over the energy interval [0,0.1] and 5000 realizations. The slope $D_{2}$ is increased with increasing disorder while the points for $IPR$ for small disorder lie above the corresponding points for larger disorder, in agreement with  Fig. 7.}
          \label{Fig10}
\end{figure}

\begin{figure}
          \centering
                   \includegraphics[width=0.9\columnwidth,clip=true]{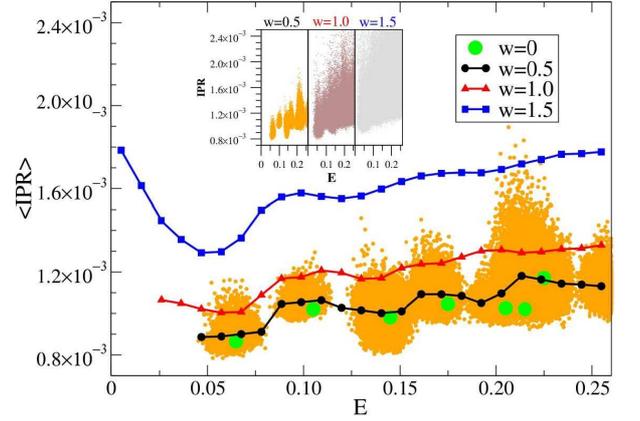}
          \caption{$\langle IPR \rangle$ for a hexagonal flake with armchair edges of size $L=20.2$(2382 sites) for different disorder strengths and 5000 realizations along with $IPR$ in the background and the inset including the $w=0$ case. Overall, $IPR$ increases with increasing disorder like in the case of the trigonal armchair flake. This is the behavior observed in the diffusive regime of normal disordered systems.}
          \label{Fig11}
\end{figure}

\begin{figure}
\centering
\includegraphics[width=0.5\columnwidth,angle=270]{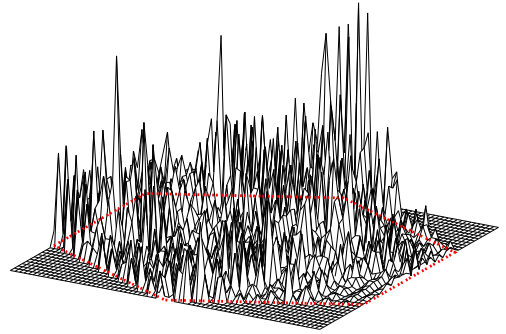}
\caption{The wavefunction probability for a hexagonal armchair flake consisting of $2382$ sites($L=20.2$) for strength of disorder $w=1.5$ at energy $E\sim0.07$. The amplitude is spread on the whole lattice, fluctuating randomly,a characteristic example of a diffusive wavefunction.
}
\label{Fig12}
\end{figure}

\begin{figure}
          \centering
                   \includegraphics[width=0.9\columnwidth,clip=true]{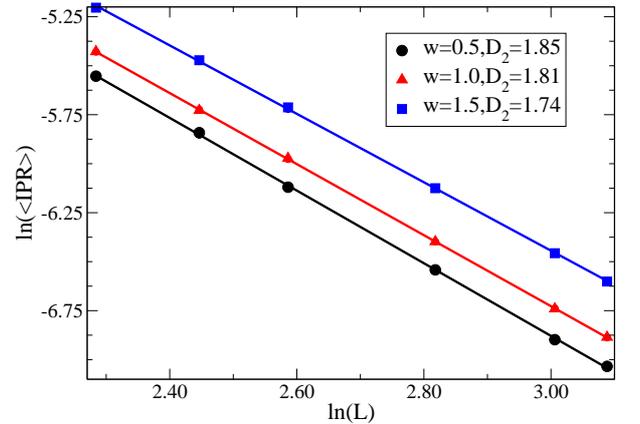}
          \caption{The scaling of $IPR$ for a hexagonal flake with armchair edges for  disorder strengths $w=0.5, 1, 1.5$ averaged in the energy interval $[0,0.2]$ and $5000$ realizations. $D_{2}$ behaves conventionally, decreasing with increasing disorder.}
          \label{Fig13}
\end{figure}

To finalize our study, we consider the case of hexagonal flakes with armchair edges shown in Fig. 1(d). In this case, there are no edge states in the limit of zero disorder, in contrast to the flakes with zigzag edges. In Fig. 11, we show $\langle IPR\rangle$ for a hexagonal flake with armchair edges with $L=20.2$ consisting of 2382 sites for disorder strengths $w=0,0.5,1.5$ and 5000 realizations along with the corresponding $IPR$ values in the background and the inset. Overall $IPR$ increases as the disorder is increased like in the case of the trigonal shape with armchair edges. This is compatible with the behavior observed in normal disordered systems like a square lattice or a linear chain, as we have already pointed out. Also in Fig. 13, we observe that $D_{2}$ is clearly decreased. In Fig. 12, the corresponding wavefunction has the characteristic diffusive form with randomly fluctuating amplitude covering the whole lattice, compatible with the non integer values of $D_{2}$, close to two. So, the hexagonal flakes with armchair edges at the presence of short range disorder, do not exhibit the abnormal behavior we encountered in flakes with zigzag edges which provide edge states.

\section{Discussion and Conclusions}
\label{disc_and_concl}
In this paper, we have presented numerical results for graphene flakes with short-range disorder that show a highly abnormal behavior for the localization properties of the wavefunctions when edge states are present. We observe a decrease of the Inverse participatio ratio $\langle IPR\rangle$ and an increase of the fractal dimension $D_{2}$  with increasing disorder, implying that the wavefunctions become roughly less localized as the disorder is increased. We argue that the underlying mechanism that causes this behavior is the interplay between the destructive interference mechanism that produces edge states(concentrated wavefunctions at the borders) and the diffusive interference mechanism, known to prevail in 2d mesoscopic systems with short range disorder for scales below the localization length. We have verified this behavior through the study of trigonal and hexagonal graphene flakes with zigzag  edges where edge states are present. The abnormal behavior is absent for flakes with armchair edges which do not result in edge states. Moreover the edge states survive for weak disorder. On the other hand, for sufficiently strong disorder the edge state mechanism is completely suppressed by the destructive interference mechanism of the short-range disorder(Anderson localization) resulting in localization of the wavefunctions in the bulk of the flakes instead of the edges. The abnormal behavior we obtained exists in the intermediate regime between the weak and strong disorder limit.

\begin{figure}
          \centering
         \includegraphics[width=0.9\columnwidth,clip=true]{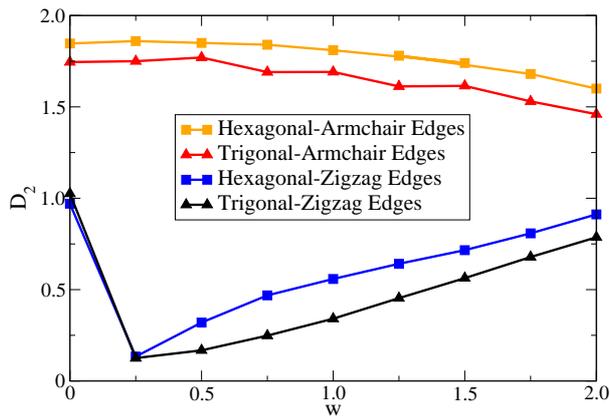}
          \caption{The fractal dimension $D_{2}$ vs. the disorder strength w for trigonal and hexagonal flakes with zigzag and armchair edges. For the flakes with zigzag edges and $w>$0.25, $D_{2}$ is increased with increasing disorder in contrast to the flakes with armchair edges for which $D_{2}$ behaves conventionally, decreasing with the disorder. This means that when edge states are present the volume of the corresponding wavefunctions is in average increasing, they become less localized with increasing disorder as we have shown through the analysis of $IPR$.  At the limit of zero disorder $(w=0)$ $D_{2}=1$ for both the zigzag triangle and hexagon since both flakes exhibit zero energy edge states extended along the flake's border. For $w=0-0.25$ localization of these edge states results in a steep decrease of $D_{2}$, followed by the abnormal behavior we described.}
        \label{Fig14}
\end{figure}

So, we have shown that when edge states are present, the consideration of interference effects in graphene systems with short-range disorder is very important and leads to unexpected behavior. In our work we concentrated in the diffusive regime while in future studies, we also intend to investigate the localized regime. Additionally, we would like to extend our analysis in order to include effects like the magnetization of the edges\cite{voznyy,oleg} in graphene systems with disorder or to investigate the connection with topological insulators which has been shown to carry resemblance to graphene\cite{buttiker1,buttiker2}, due to the edge states mimicking the topological property of the electron current flowing from the boundary surfaces in these materials. We hope that our work will motivate further experimental investigation of the edge states in graphene systems, and their impact on the electronic properties.

\begin{acknowledgement}
We would like to thank D. Katsanos, V. A. Gopar and S. N. Evangelou for useful discussions and careful reading of the manuscript.
We also acknowledge the computer resources and assistance provided by the Institute for Biocomputation and Physics of Complex Systems (BIFI) of the University of Zaragoza.
\end{acknowledgement}
*Present address:Department of Electrophysics, National Chiao Tung University,
1001 University Road, Hsinchu 30010, Taiwan


\begin{thebibliography}{50}



\bibitem{Geim} Geim, A. K. and Novoselov, K. S, {\it Nature Materials} {\bf6},183-191 (2007).

\bibitem{Novoselov}K.S.Novoselov et al., {\it Science} {\bf306}, 666 (2004).

\bibitem{fujita} K. Nakada, M. Fujita G. Dresselhaus, M. S. Dresselhaus, {\it Phys. Rev. B} {\bf54}, 17954 (1996).

\bibitem{wakaplus}Wakabayashi, K., M. Fujita, H. Ajiki, and M. Sigrist, {\it Phys. Rev. B} {\bf59}, 8271 (1999).
\bibitem{akhemerov} A. Akhmerov and CWJ Beenakker, {\it Phys. Rev. B} {\bf77}, 085423 (2008)
\bibitem{Li} X. Li, X. Wang, L. Zhang, S. Lee, and H. Dai, {\it Science} {\bf319}, 1229 (2008).

\bibitem{Jiao} Liying Jiao, Li Zhang, Xinran Wang, Georgi Diankov,  Hongjie Dai, {\it Nature} {\bf458} (7240): 877-80 (2009).

\bibitem{Brumfiel} G. Brumfiel {\it Nature}, {\bf458} (2009), p. 390.

\bibitem{Fuhrer} Michael S. Fuhrer, {\it Nature Materials},{\bf9},611-612 (2010).

\bibitem{taoplus} C. Tao, L. Jiao, O. V. Yazyev, Y.-C. Chen, J. Feng, X. Zhang, R. B. Capaz, J. M. Tour, A. Zettl, S. G. Louie, H. Dai, and M. F. Crommie, {\it Nat. Phys.} {\bf7}, 616 (2011).

\bibitem{yamamoto}Yamamoto, T.; Noguchi, T.; Watanabe, K. {\it Phys. Rev. B}  {\bf74} (12), 121409 (2006).

\bibitem{ezawa} M. Ezawa, {\it Phys. Rev. B } {\bf76}, 245415 (1-6) (2007)

\bibitem{wang} W. L. Wang, S. Meng, and E. Kaxiras, Nano Lett. {\bf8}, 241 (2008)

\bibitem{heiskanen1}H. P. Heiskanen and M. Manninen, J. Akola, {\it New J. Phys.} {\bf10}, 103015 (2008).

\bibitem{heiskanen2} Akola J, Heiskanen H P and Manninen M., {\it Phys. Rev. B} {\bf77}, 193410 (2008).


\bibitem{ponomarenko} L. A. Ponomarenko et al., {\it Science} {\bf320}, 356 (2008).

\bibitem{wu}Wu J., Pisula W.,  M\"{u}llen K. {\it Chemical Reviews}, Vol. {\bf107}(3) (October, 2007) pp. 718-747

\bibitem{zu} Zhi L., M\"{u}llens K. {\it Journal of Materials Chemistry}, Vol. {\bf18} (February, 2008) pp. 1472-1484

\bibitem{Guttinker}J. Guttinger, C. Stampfer, S. Hellmller, F. Molitor, T. Ihn, and K. Ensslin, {\it Appl. Phys. Lett.} {\bf93}, 212102 (2008).

\bibitem{Schnez} S. Schnez, F. Molitor, C. Stampfer, J. Guettinger, I. Shorubalko, T. Ihn, and K. Ensslin,{\it Appl. Phys. Lett.}
{\bf94}, 012107 (2009).
\bibitem{neto}Castro Neto, A. H., F. Guinea, N. M. R. Peres, K. S. Novoselov, and A. K. Geim , {\it Rev. Mod. Phys.} {\bf81}, 109
(2009).

\bibitem{lewenkopf}Mucciolo, E. R., and C. H. Lewenkopf, {\it J. Phys.: Condens. Matter} {\bf22}, 273201 (2010).

\bibitem{voznyy}  O. Voznyy, A. D. G\"{u}cl\"{u}, P. Potasz, and P. Hawrylak, {\it Phys. Rev. B} {\bf83}, 165417 (2011).
\bibitem{oleg}Oleg V. Yazyev 2010 {\it Rep. Prog. Phys.} {\bf73} 05650

\bibitem{kuhl}S. Barkhofen, M. Bellec, U. Kuhl, F. Mortessagne,  {\it Phys. Rev. B}, {\bf87}, 035101 (2013)

\bibitem{beenacker} A. Rycerz, J. Tworzydlo, and C. W. J. Beenakker, {\it Nat. Phys.} {\bf3}, 172 (2007) [CAS].

\bibitem{ando1}  Ando, T., and T. Nakanishi, 1998, {\it J. Phys. Soc. Jpn.} {\bf67}, 2857.

\bibitem{ando2} T. Ando, NPG {\it Asia Mater}, {\bf1} (1) (2009), pp. 17–21.

\bibitem{wakapcc1} K. Wakabayashi, Y. Takane, and M. Sigrist, {\it Phys. Rev. Lett.} {\bf99}, 036601 (2007).

\bibitem{wakapcc2}  K. Wakabayashi, Y. Takane, M. Yamamoto, and M. Sigrist, {\it Carbon} {\bf47}, 124 (2009)[CAS].

\bibitem{anderson1} P. W. Anderson, {\it Phys. Rev.} {\bf109}, 1492 (1958).

\bibitem{anderson2} P. W. Anderson, D. J. Thouless, E. Abrahams, and D. S. Fisher, {\it Phys. Rev. B} {\bf22}, 3519 (1980).

\bibitem{nanjing}S. J. Xiong and Y. Xiong,{\it Phys. Rev. B} {\bf76}, 214204 (2007).

\bibitem{romer1}C. González-Santander, F. Domínguez-Adame, M. Hilke, R. A. R\"{o}mer,
{\it EPL} {\bf104}, 17012-6 (2013).

\bibitem{amanatidis}I. Amanatidis and S. N. Evangelou, {\it Phys. Rev. B} {\bf79}, 205420 (2009).

\bibitem{huang}Liang Huang, Ying-Cheng Lai, and Celso Grebogi {\it Phys. Rev. E} {\bf81}, 055203 (2010).

\bibitem{akhmerov} M. Wimmer, A. R. Akhmerov, and F. Guinea, {\it Phys. Rev. B} {\bf82}, 045409 (2010).

\bibitem{rycerz}Adam Rycerz, Phys. Rev. B {\bf85}, 245424 (2012).

\bibitem{amanatidis1}H. Amanatidis, I. Kleftogiannis, D.E. Katsanos and S.N Evangelou, {\it cond-mat.mes-hall} arXiv:1302.2470 (2013)


\bibitem{weinmann}Dietmar Weinmann, Sigmund Kohler, Gert-Ludwig Ingold,and Peter Hanggi Ann. Phys. (Leipzig) (1999),
{\it Spec. Issue}, SI-277-SI-280.

\bibitem{falko} V.I. Falko and K.B. Efetov, Europhys. Lett. 32, 627 (1995); {\it Phys. Rev. B} {\bf52}, 17413 (1995).



\bibitem{Hentschel}H. Hentschel and I. Procaccia, {\it Physica} (Amsterdam) {\bf8D}, 435 (1983).

\bibitem{spiros}S. N. Evangelou,{\it J. Phys. A} {\bf23}, L317 (1990).

\bibitem{chamon}Claudio de C. Chamon, Christopher Mudry, and Xiao-Gang Wen  {\it Phys. Rev. Lett.}
{\bf77}, 4194 (1996).

\bibitem{castillo} Castillo H E, de C Chamon C, Fradkin E, Goldbart P M and Mudry C {\it Phys. Rev. B} {\bf56},
10668 (1997).

\bibitem{subramaniam}Subramaniam, A. R., I. A. Gruzberg, A. W. W. Ludwig, F.Evers, A. Mildenberger
and A. D. Mirlin {\it Phys. Rev. Lett.} {\bf 96}, 126802 (2006).

\bibitem{evers} F. Evers and A. D. Mirlin, {\it Rev. Mod. Phys.}  {\bf 80}, 1355 (2008).

\bibitem{romer}A. Rodriguez, L. J. Vasquez, K. Slevin, R. A. R\"{o}mer {\it Phys. Rev. B} {\bf84}, 134209 (2011).

\bibitem{kleftogiannis}I. Kleftogiannis, S.N. Evangelou, {\it cond-mat.mes-hall} arXiv:1304.5968 (2013)



\bibitem{buttiker1}J. Li, I. Martin, M. B\"{u}ttiker, and A. F. Morpurgo,{\it Nature Phys.} {\bf7}, 38 (2011).

\bibitem{buttiker2}B\"{u}ttiker  M., 2009, {\it Science} {\bf325}, 278.


\end{thebibliography}
\end{document}